\documentclass[conference,compsoc]{IEEEtran}
\usepackage{calc}
\usepackage{hyperref}
\usepackage{xspace}
\usepackage{xcolor}
\usepackage{array}
\usepackage{setspace}
\usepackage{multirow}
\usepackage{booktabs}
\usepackage{enumitem}
\usepackage{listings}
\usepackage{algorithm}
\usepackage{algorithmic}

\usepackage{etoolbox}  
\makeatletter
\patchcmd{\algorithmic}{\addtolength{\ALC@tlm}{\leftmargin} }{\addtolength{\ALC@tlm}{\leftmargin}}{}{}
\makeatother

\newcommand\algorithmicprocedure{\textbf{function}}
\newcommand{\algorithmicendprocedure}{\algorithmicend\ \algorithmicprocedure}
\makeatletter
\newcommand\PROCEDURE[3][default]{%
  \ALC@it
  \algorithmicprocedure\ \textsc{#2}(#3)%
  \ALC@com{#1}%
  \begin{ALC@prc}%
}
\newcommand\ENDPROCEDURE{%
  \end{ALC@prc}%
  \ifthenelse{\boolean{ALC@noend}}{}{%
    \ALC@it\algorithmicendprocedure
  }%
}
\newenvironment{ALC@prc}{\begin{ALC@g}}{\end{ALC@g}}
\makeatother

\newcommand{\SWITCH}[1]{\STATE \textbf{switch} #1}
\newcommand{\ENDSWITCH}{\STATE \textbf{end switch}}
\newcommand{\CASE}[1]{\STATE \textbf{case} #1\textbf{:} \begin{ALC@g}}
\newcommand{\ENDCASE}{\end{ALC@g}}
\newcommand{\CASELINE}[1]{\STATE \textbf{case} #1\textbf{:} }
\newcommand{\DEFAULT}{\STATE \textbf{default:} \begin{ALC@g}}
\newcommand{\ENDDEFAULT}{\end{ALC@g}}
\newcommand{\DEFAULTLINE}[1]{\STATE \textbf{default:} }

\definecolor{darkpastelgreen}{rgb}{0.01, 0.75, 0.24}
\definecolor{bistre}{rgb}{0.24, 0.17, 0.12}
\definecolor{lightgray}{rgb}{.9,.9,.9}
\definecolor{darkgray}{rgb}{.4,.4,.4}
\definecolor{purple}{rgb}{0.65, 0.12, 0.82}
\definecolor{cream}{rgb}{1.0, 0.99, 0.82}

\newcommand{\HilightGray}{\makebox[0pt][l]{\color{lightgray}\rule[-\dp\strutbox]{0.47\textwidth}{\baselineskip}}}
\newcommand{\HilightYellow}{\makebox[0pt][l]{\color{cream}\rule[-\dp\strutbox]{0.47\textwidth}{\baselineskip}}}

\lstdefinelanguage{JavaScript}{
  morekeywords=[1]{break, case, catch, continue, debugger, default, delete, do, else, false, finally, for, function, if, in, instanceof, new, null, return, switch, this, throw, true, try, typeof, var, void, while, with, WebAssembly, let, const, Debugger, wasmTextToBinary, Reflect, Date},
  morecomment=[l]{//},
  morecomment=[s]{/*}{*/},
  morestring=[b]',
  morestring=[b]",
  morekeywords=[2]{class, boolean, throw, implements, this},
  keywordstyle=[1]\color{blue}\bfseries,
  keywordstyle=[2]\color{darkgray}\bfseries,
  identifierstyle=\color{black},
  commentstyle=\color{darkpastelgreen}\ttfamily,
  stringstyle=\color{red}\ttfamily,
  sensitive=true
}

\lstdefinestyle{js-script-size}{
	language=JavaScript,
	extendedchars=true,
	basicstyle=\scriptsize\ttfamily,
	showstringspaces=false,
	showspaces=false,
	numbers=left,
    xleftmargin=10pt,
	numberstyle=\scriptsize,
	numbersep=6pt,
	tabsize=2,
	breaklines=true,
	showtabs=false,
	captionpos=b,
	frame=none,
}

\lstdefinestyle{js-small-size}{
	language=JavaScript,
	extendedchars=true,
	basicstyle=\small\ttfamily,
	showstringspaces=false,
	showspaces=false,
	numbers=left,
    xleftmargin=10pt,
	numberstyle=\small,
	numbersep=6pt,
	tabsize=2,
	breaklines=true,
	showtabs=false,
	captionpos=b,
	frame=none,
}

\lstdefinelanguage{WebAssembly}{
    sensitive=true,
    alsoletter={.},
	morekeywords=[1]{module, func, export, memory, global}, 
	keywordstyle=[1]\color{blue}\bfseries,
    morekeywords=[2]{i32.const, call, loop, block, end, drop, if, else, f64.const, local.get, f64.copysign},	
	keywordstyle=[2]\color{bistre}\bfseries,
	morekeywords=[3]{funcref, externref, i32, i64, f32, f64, v128},	
	keywordstyle=[3]\color{teal}\bfseries,
	identifierstyle=\color{black},
    morecomment=[l]{;;},
	morecomment=[s]{(;}{;)},
	commentstyle=\color{darkpastelgreen}\ttfamily,
	stringstyle=\color{red}\ttfamily,
	morestring=[b]',
	morestring=[b]"
}

\lstdefinestyle{wasm}{
	language=WebAssembly,
	extendedchars=true,
	basicstyle=\scriptsize\ttfamily,
	showstringspaces=false,
	showspaces=false,
	numbers=left,
    xleftmargin=10pt,
	numberstyle=\scriptsize,
	numbersep=6pt,
	tabsize=2,
	breaklines=true,
	showtabs=false,
	captionpos=b,
	frame=none,
}

\lstdefinestyle{context}{
	extendedchars=true,
	basicstyle=\small\ttfamily,
	showstringspaces=false,
	showspaces=false,
	numbers=left,
    xleftmargin=10pt,
	numberstyle=\small,
	numbersep=6pt,
	tabsize=2,
	breaklines=true,
	showtabs=false,
	captionpos=b,
	frame=none,
}

\newcommand{\coolname}{\textsc{Weaver}\xspace}

\ifCLASSOPTIONcompsoc
  \usepackage[nocompress]{cite}
\else
  \usepackage{cite}
\fi
\ifCLASSINFOpdf
  \usepackage[pdftex]{graphicx}
\else
\fi
\usepackage{amsmath}
\ifCLASSOPTIONcompsoc
  \usepackage[caption=false,font=footnotesize,labelfont=sf,textfont=sf]{subfig}
\else
  \usepackage[caption=false,font=footnotesize]{subfig}
\fi
\usepackage{xurl}

\begin{document}
%
\title{\coolname: Fuzzing JavaScript Engines at the JavaScript-WebAssembly Boundary}

\author{\IEEEauthorblockN{Lingming Zhang\IEEEauthorrefmark{1}\IEEEauthorrefmark{2},
Binbin Zhao\IEEEauthorrefmark{1},
Puzhuo Liu\IEEEauthorrefmark{2},
Qinge Xie\IEEEauthorrefmark{3},
Peng Di\IEEEauthorrefmark{2},
Jianhai Chen\IEEEauthorrefmark{1},
Shouling Ji\IEEEauthorrefmark{1}
}
\IEEEauthorblockA{\IEEEauthorrefmark{1}Zhejiang University, China\\
\{lingming.zhang, binbinz, chenjh919, sji\}@zju.edu.cn}
\IEEEauthorblockA{\IEEEauthorrefmark{2}Ant Group, China\\
\{liupuzhuo.lpz, dipeng.dp\}@antgroup.com}
\IEEEauthorblockA{\IEEEauthorrefmark{3}Georgia Institute of Technology, USA\\
qxie47@gatech.edu}

}


\maketitle

\begin{abstract}

The security of modern JavaScript (JS) engines is critical since they provide the primary defense mechanism for executing untrusted code on the web. The recent integration of WebAssembly (Wasm) has transformed these engines into complex polyglot environments, creating a novel attack surface at the JS-Wasm interaction boundary due to the distinct type systems and memory models of two languages. This boundary remains largely underexplored, as previous works mainly focus on testing JS and Wasm as two isolated entities rather than investigating the security implications of their cross-language interactions.

This paper proposes \coolname, an effective greybox fuzzing framework specifically tailored to uncover vulnerabilities at the JS-Wasm boundary. To comply with the language constraints, \coolname uses a type-aware generation strategy, meticulously maintaining the dual-type representation for every generated variables. This allows fuzzer to validly utilize variables across the language boundary. Besides, \coolname leverages the UCB-1 algorithm to intelligently schedule mutators and generators to maximize the discovery of new code paths.

We have implemented and evaluated \coolname on three JS engines. The results indicate that \coolname achieves superior code coverage compared to state-of-the-art fuzzers. Moreover, \coolname has uncovered two new bugs in the latest versions of these engines, one of which is considered high severity and set to highest priority, demonstrating the practicality of \coolname.

\end{abstract}


%
\IEEEpeerreviewmaketitle

\section{Introduction}\label{sec:intro}

JavaScript (JS) is a cornerstone of the modern Internet, enabling dynamic and interactive user experiences for over 98.8\% of all the websites~\cite{w3techs-report}. To execute JS code, all major web browsers and server-side runtimes (e.g., Node.js) rely on high-performance JS engines, such as SpiderMonkey~\cite{spidermonkey}, V8~\cite{v8}, and JavaScriptCore~\cite{javascriptcore}. One core responsibility of these engines is to provide a secure and robust environment for executing untrusted code from arbitrary sources. Thus, the security of the JS engine itself is of vital importance. Any vulnerability occurred at the engine level can break the last line of defense and result in severe consequences, such as remote code execution~\cite{bug1452137-exploit, CVE-2024-12695}, posing a critical threat to user security.

In the past few years, the introduction of WebAssembly (Wasm)~\cite{wasm-paper} has fundamentally altered the internal execution environment of JS engines. Wasm is a safe, portable binary instruction set and serves as the compilation target for high-level programming languages like C++. The integration of Wasm into JS engines enables traditional high-performance applications to be run on the web yet simultaneously turns JS engine into a complex \textbf{polyglot environment}, where JS code and Wasm module coexist and could perform efficient inter-operations. This sheds light on a novel attack surface, namely, the boundary between two languages with distinct type systems and memory models.

To date, little attention has been paid to the vulnerability discovery at the language boundary. Previous research works on JS engines usually treat JS and Wasm as isolated entities. They either focus on detecting memory bugs or logical flaws in the JS interpreter and Just-in-Time (JIT) compiler~\cite{CodeAlchemist,OptFuzz,JITPicking, Dumpling}, or devote to detecting issues in Wasm subsystem's parsing, compilation, and execution phases~\cite{RGFuzz,Waltzz}. The security issues raised during the cross-language interaction remains an underexplored area.

In this paper, we attempt to leverage fuzzing to uncover vulnerabilities at the JS-Wasm interaction boundary. Fuzzing is considered one of the most effective techniques for finding bugs, which generates a large amount of test cases and feeds them to the target to detect any abnormal behaviors. Unlike static analysis techniques, fuzzing scales smoothly to larger programs, especially under the context of JS engine, and is mostly free from false positives. However, building a fuzzer that can effectively detect bugs in JS-Wasm interactions still faces the following challenges:

\noindent\textbf{Challenge \uppercase\expandafter{\romannumeral1}: Compliance with dual-language validation rules.} The semantic correctness of test cases has constantly been a major concern in the field of domain-specific fuzzing and is extremely important within JS engine research, given that the prevalence of vulnerabilities is shifting from parser and interpreter to optimizing compiler~\cite{Die}. The difficulty lies in the fact that JS and Wasm are two distinct languages with different validation rules, which means the interaction across two languages must respect each language's semantic requirements. This demands the fuzzer to possess a deep and comprehensive understanding of both languages' semantics and the specific rules governing the interaction interface to produce meaningful test cases.

\noindent\textbf{Challenge \uppercase\expandafter{\romannumeral2}: Effective navigation of combinatorial input space.} The generation of inputs involves the JS and Wasm part, both of which encompass rich feature sets. For instance, JS itself contains over 50 standard built-in objects featuring varied purposes. The combinatorial explosion of the potential interaction space necessitates that a fuzzer possess the capacity for efficiently exploring complex and rare JS-Wasm interactions.

In regard to these challenges, we propose \coolname, an effective greybox fuzzing framework specifically tailored for JS-Wasm interaction. To address \textbf{Challenge \uppercase\expandafter{\romannumeral1}}, we adopt the type-aware test case generation, which has proven useful in previous studies~\cite{Fuzzilli}. To satisfy the semantic constraints of both interacting parties, \coolname meticulously maintains a dual-type representation for every generated object, signifying its JS and Wasm types. Building on this foundation, the generation process proceeds in a unified manner: it directly retrieves objects matching the requisite JS/Wasm types from the surrounding context, without tracking whether the object comes from the JS or Wasm domain, thereby allowing Wasm objects to be correctly used in the JS context and vice versa. To tackle \textbf{Challenge \uppercase\expandafter{\romannumeral2}}, we model the fuzzing campaign as a multi-armed bandit problem~\cite{ecofuzz} and leverage the UCB-1 algorithm to select generators and mutators that maximize the possibility of discovering new code paths.

We implement \coolname and comprehensively evaluate it on a set of representative JS engines from major vendors, including JavaScriptCore from Apple, V8 from Google, and SpiderMonkey from Mozilla. Compared to state-of-the-art fuzzers such like Fuzzilli~\cite{Fuzzilli}, \coolname achieves superior performance on both line and branch coverage. Moreover, \coolname has found two previously unknown bugs in the latest versions of JavaScriptCore and SpiderMonkey. Notably, the bug found in SpiderMonkey is deemed high-severity by developers as it could enable arbitrary code execution. This highlights the severity of bugs identified by~\coolname.

\noindent\textbf{Contributions.} In summary, our main contributions include:
\begin{itemize}
    \item We propose \coolname, a novel approach for fuzzing the JS-Wasm interaction boundary, which has remained a largely underexplored area. \coolname utilizes the dual-type representation and UCB-1 algorithm to generate test cases that satisfy language constraints while simultaneously exploring diverse interactions.
    \item We conduct thorough experiments which show that \coolname outperforms state-of-the-art baselines such as Fuzzilli in terms of code coverage. Besides, \coolname's long-term fuzzing campaign has led to the discovery of two new bugs, one of which is set to high severity and top priority, demonstrating the efficacy of \coolname on finding real-world bugs. In total, we have been awarded \$1K in bug bounties for our findings.
\end{itemize}
\section{Background}

In this section, we first introduce the fundamental characteristics of JS and Wasm and their interaction mechanisms, followed by a motivating example to articulate the premise of this work.

\subsection{JS-Wasm Interaction}\label{sec:background-interaction}

Modern JS engines such as SpiderMonkey can be viewed as an extremely complex trusted computing base whose core responsibility is to execute two types of code formats which share intrinsically distinct design paradigms:

\begin{enumerate}
    \item JS is a high-level programming language with dynamic typing, which indicates a variable's type is determined at runtime and could be changed at any point during execution. For instance, \texttt{dynamicVar} in Figure~\ref{code:js-sample} is declared as a \texttt{number} type at start and later reassigned as the \texttt{string} type without raising any runtime error. Besides, JS adopts garbage collection to automatically trace and reclaim memory objects that are no longer referenced.
    \item Wasm is a low-level binary instruction set designed as a compilation target for languages like C++ and Rust. In contrast to JS, Wasm is statically typed and its code must pass the validation phase to ensure type safety. For example, the \texttt{global} in Figure~\ref{code:wasm-sample} is declared as an \texttt{i32} type, which remains the same across execution. Moreover, Wasm allows explicit linear memories to be declared to support the manual memory management practice.
\end{enumerate}

While JS and Wasm are essentially different, JS engines still provide mechanisms for them to interact bidirectionally, thereby allowing both to function more effectively. Within the JS context, there are mainly seven distinct Wasm object types available for use: \textcircled{1} \textbf{Memory} represents a randomly accessible linear array that holds raw bytes; \textcircled{2} \textbf{Table} is used to store references which could either come from Wasm or JS; \textcircled{3} \textbf{Global} stands for global variable that can be declared as mutable or immutable and further provided initialization value; \textcircled{4} \textbf{Tag} defines the shape of an exception that can be thrown in Wasm code; \textcircled{5} \textbf{Function} denotes a callable Wasm function which comprises function signature, types of local variables, and function body; \textcircled{6} \textbf{Module} is a stateless Wasm program that collects definitions of the aforementioned five Wasm objects and might also declares those objects as either imported from or exported to the JS context; \textcircled{7} \textbf{Instance} is a stateful, executable instance of a Wasm module, providing the JS environment with access to all its exported objects. According to the Wasm-JS Interface specification~\cite{wasm-js-api}, the first four Wasm object types can be instantiated directly via the JS APIs or exported through a Wasm instance. Functions are exclusively exported by Wasm instances, while the final two objects are solely constructed via the JS APIs. Figure~\ref{code:interact-sample} presents an illustrating example where \texttt{jsmem} is constructed through the \texttt{WebAssembly.Memory} API and \texttt{wasmmem} is exported to the JS environment from the Wasm instance \texttt{inst}. Ideally, all Wasm objects can be viewed as \texttt{object} types under JS with different properties and methods, which enables interactions with different JS built-ins. For instance, memory object contains a \texttt{buffer} property which points to the underlying memory buffer. Since \texttt{buffer} property is of type \texttt{ArrayBuffer}, it can be further utilized to create a \texttt{DataView} as shown in Figure~\ref{code:interact-sample}. From the other side, Wasm accepts JS objects through the import declaration as long as the object matches the predefined Wasm type. For example, the JS function \texttt{greet} in Figure~\ref{code:interact-sample} can be treated as a Wasm function that takes no parameters and yields no return values, allowing it to be imported and invoked under Wasm context.

As can be seen from previous discussion, the interaction between JS and Wasm introduces a new layer of complexity to JS engines, where the interleaving of cross-language data and control flows becomes commonplace. Failure to handle the cross-language interaction correctly could introduce new security vulnerabilities into the engine and we will discuss one such example in Section~\ref{sec:motivating-example}.

\newbox\jsbox
\begin{lrbox}{\jsbox}
\begin{lstlisting}[
    style=js-script-size,
    linewidth=0.48\textwidth
]
let dynamicVar = 100;
// dynamically typed.
dynamicVar = "Hello";
dynamicVar = {key: "value"};
// garbage collection.
let data = new Array(100000);
data.fill("Test");
\end{lstlisting}
\end{lrbox}

\newbox\wasmbox
\begin{lrbox}{\wasmbox}
\begin{lstlisting}[
    style=wasm,
    linewidth=0.48\textwidth
]
(module
  ;; explicit memory.
  (memory 1 10)
  ;; statically typed.
  (global i32)
  (func (result i64))
)
\end{lstlisting}
\end{lrbox}

\newbox\interactionbox
\begin{lrbox}{\interactionbox}
\begin{lstlisting}[
    style=js-script-size,
    linewidth=0.48\textwidth
]
function greet() {
  console.log("Hello World!");
}
let jsmem = new WebAssembly.Memory({initial:1});
// (module
//   (import "js" "greet" (func))
//   (memory (;0;) 1 5)
//   (export "fun" (func 1))
//   (export "mem" (memory 0))
//   (func (param i32) (result i32)
//     call 0
//     local.get 0
//   )
// )
let mod = new WebAssembly.Module(new Uint8Array(...));
let inst = 
    new WebAssembly.Instance(mod, {js:{greet:greet}});
let wasmmem = inst.exports['mem'];
let wasmfun = inst.exports['fun'];
// a possible data-flow interaction.
let view = new DataView(wasmmem.buffer);
// a possible control-flow interaction.
wasmfun(0);
\end{lstlisting}
\end{lrbox}

\begin{figure}[t]
    \centering
    \subfloat[JS code sample.\label{code:js-sample}]{\usebox\jsbox}
    \hspace{0.5cm}
    \subfloat[Wasm code sample.\label{code:wasm-sample}]{\usebox\wasmbox}\\
    \subfloat[JS-Wasm interaction code sample.\label{code:interact-sample}]{\usebox\interactionbox}
    \caption{The code examples illustrating key characteristics of JS and Wasm along with their interaction.}
\end{figure}

\subsection{Motivating Example}\label{sec:motivating-example}

Here, we center around a previously identified vulnerability to present the motivation of this work, discuss why prior research is inadequate for discovering such flaws, and demonstrate the challenges in detecting this class of bugs.

Figure~\ref{code:motivating-example} illustrates a real-world vulnerability discovered in SpiderMonkey which involves three steps to trigger. First, Line 17-19 employ a for-loop to repeatedly execute function \texttt{b}, which activates JS engine's JIT compilation to compile function \texttt{b} into more efficient machine code. This lays the foundation for following steps. Next, the function \texttt{b} defines, compiles, and instantiates a Wasm module. The crux of this step lies in the Wasm module importing and invoking a JS function \texttt{c}. When we call the exported Wasm function \texttt{test} at Line 14, this effectively creates a JS$\rightarrow$Wasm$\rightarrow$JS cross-language control flows, which sets up the scene for the final bug triggering. At last, the function \texttt{c} utilizes the Debugger interface to inspect call frames of previously executed code. However, the Wasm frame debugger in SpiderMonkey fails to correctly handle or skip non-Wasm frames when iterating through stack frames, which results in a corrupted state and eventually triggers an assertion failure.

\begin{figure}[t]
    \centering
\begin{lstlisting}[
        style=js-script-size,
        escapechar=\%,
        linewidth=0.48\textwidth
    ]
function c() {
%\HilightYellow%  var dbg =
%\HilightYellow%    newGlobal({newCompartment: true}).Debugger(this);
%\HilightYellow%  dbg.getNewestFrame().older.offset;
}
function b() {
%\HilightGray%  var bin = wasmTextToBinary(`
%\HilightGray%    (module
%\HilightGray%      (import "m" "f" (func $f))
%\HilightGray%      (func (export "test") call $f))
%\HilightGray%    `);
%\HilightGray%  var mod = new WebAssembly.Module(bin);
%\HilightGray%  var inst = new WebAssembly.Instance(mod, {m: {f: c}});
%\HilightGray%  inst.exports.test();
%\HilightGray%  inst.exports.test();
}
%\HilightYellow%for (var i = 0; i < 5; i++) {
%\HilightYellow%  b();
%\HilightYellow%}
\end{lstlisting}
    \caption{Proof-of-Concept of the SpiderMonkey Bug-1900740.}
    \label{code:motivating-example}
\end{figure}

From the analysis above, it is apparent that this security flaw is not purely a JS or Wasm bug, it emerges only from the interaction of two distinct languages. Therefore, this bug effectively eludes existing state-of-the-art fuzzers which aim to find bugs introduced by single language. Specifically, the JS-centric fuzzers such as DIE~\cite{Die} and OptFuzz~\cite{OptFuzz} utilize sophisticated test case generation and mutation strategies or customized feedback to find bugs in JS interpreter and JIT compiler. It is possible for them to craft the Debugger API calls and the JIT triggers such as the ones illustrated in the yellow shaded region of Figure~\ref{code:motivating-example}. However, they are mostly agnostic to the generation of Wasm-related components and, consequently, incapable of synthesizing the requisite cross-language interactions. Wasm-centric fuzzers~\cite{Waltzz,RGFuzz}, on the other hand, excel at generating complex and valid Wasm modules to test the Wasm parser, validator, and runtime, but they generally rely on a simple, static test harness template to execute the generated Wasm modules under JS engines. For instance, Waltzz~\cite{Waltzz} is configured to produce modules that require no imports and export one function called \texttt{main}. The sole purpose of the test harness applied in Waltzz is to instantiate the passed-in Wasm module and call the exported Wasm function \texttt{main}, similar to the gray shaded region of Figure~\ref{code:motivating-example}. Given the dearth of JS interaction within their test harnesses, these Wasm-centric fuzzers are similarly unable to detect this vulnerability.

As a result, there still lacks a dedicated fuzzer for testing the JS-Wasm interaction boundary, which mainly motivates this work. To effectively craft such interactions, we identify two primary challenges: \textcircled{1} Test cases must comply with not only the semantic constraints enforced by each language, but also the specific rules imposed by the interaction interface. Failure to follow relevant rules can cause the JS engine to raise a runtime error and terminate prematurely, wasting the fuzzing time. For instance, the instantiation of module \texttt{mod} in Figure~\ref{code:motivating-example} requires the JS environment to provide a function taking no arguments and yielding no result as import. If we replace function \texttt{c} with a randomly selected function such as \texttt{Atomics.add}, the JS engine would immediately report an uncaught \texttt{TypeError}, blocking the fuzzer from deeper vulnerability discovery. \textcircled{2} The generation of test cases have to respect the diverse language features of both worlds. The bug shown in Figure~\ref{code:motivating-example} requires specific Debugger interface to trigger. Given that the JS portion alone has encompassed over 50 built-ins such like \texttt{Proxy} and \texttt{Promise}, effective exploration of different features becomes a practical concern for fuzzers to find bugs akin to the illustrated example. With the aforementioned challenges in mind, we demonstrate the detailed design of \coolname in Section~\ref{sec:design}.
\section{\coolname Design}\label{sec:design}

\begin{figure}
    \centering
    \includegraphics[width=\linewidth]{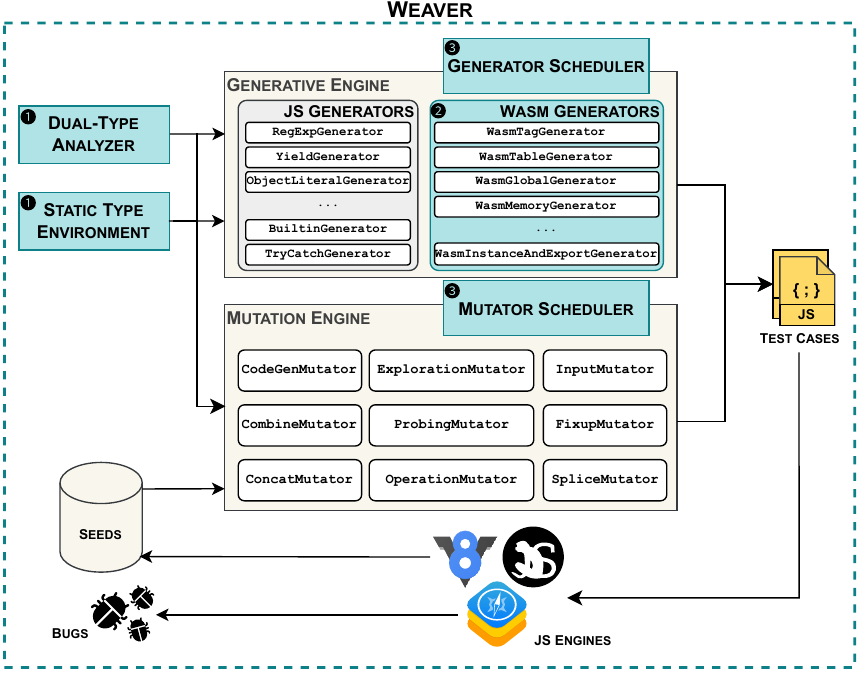}
    \caption{High-level overview of \coolname.}
    \label{fig:overview}
\end{figure}

As illustrated in Figure~\ref{fig:overview}, \coolname is a coverage-guided fuzzer dedicated to finding bugs at the JS-Wasm interaction boundary. \coolname itself builds upon the state-of-the-art JS fuzzer Fuzzilli~\cite{Fuzzilli} and comprises two primary engines. The generative engine is responsible for the corpus generation at fuzzing's initial stage, which uses the predefined generators to construct inputs. The mutation engine ignites when there are enough interesting test cases in the seed pool and applies consecutive mutations on top of randomly selected seeds to produce new test cases. In compliance with common fuzzing practices, \coolname instruments the target engine to collect coverage information, which is utilized to guide the test case preservation. Besides, any test cases crashing the engine are also retained during fuzzing for further analysis.

To enable efficient and effective interaction exploration, \coolname extends the existing Fuzzilli architecture with the following key components. First, the dual-type analyzer and static type environment deal with the type annotation of each variable created during the test case construction. To support a unified generation process, we label each variable with its corresponding JS and Wasm type, regardless of whether this variable originates from the JS part or the Wasm part. In this way, \coolname can efficiently query the context for variables matching the required JS or Wasm type, integrate them into the type-aware generation process, and consequently satisfy the semantic constraints imposed by \textbf{Challenge \uppercase\expandafter{\romannumeral1}}. Second, the Wasm generators aim to provide usable Wasm objects to the JS context and, conversely, to adopt available JS/Wasm objects within the Wasm context, which allow the fuzzer to assemble meaningful cross-language interactions. Third, to enhance fuzzer's ability to explore difficult-to-reach or rare features, \coolname proposes performance-based schedulers for both the generators and the mutators which handle the exploration–exploitation problem during fuzzing. The Wasm generators and schedulers jointly support \coolname to tackle \textbf{Challenge \uppercase\expandafter{\romannumeral2}}.

In the next, we first present a running example demonstrating the generation procedure of \coolname in Section~\ref{sec:running-example} and then discuss the detailed design of dual-type analysis, Wasm generators, and performance-based schedulers in Section~\ref{sec:dual-type-analysis}, Section~\ref{sec:wasm-generation}, and Section~\ref{sec:scheduling}, respectively.

\subsection{A Running Example}\label{sec:running-example}

In this section, we leverage a running example to clearly illustrate how \coolname generates a semantically correct test case with diverse JS-Wasm interactions involved. The input generation process generally starts by populating a number of preamble variables that require no other variable as input and further serve as inputs for other generators. For instance, \coolname may choose the generator \texttt{BuiltinGenerator} to introduce a JS built-in \texttt{Reflect}:

\begin{lstlisting}[style=js-small-size]
const v1 = Reflect;
\end{lstlisting}

The dual-type analyzer is then initiated to determine the JS and Wasm types of \texttt{v1}. Since \texttt{v1} stands for a JS built-in, the analyzer queries the static type environment to retrieve its preassigned JS type, namely, an \texttt{object(..)} type. The analyzer then infers its possible Wasm types based upon the obtained JS type which produces \texttt{\{(ref extern),(ref any),externref,anyref\}}. The first one represents a non-nullable reference to external values which are outside the control of Wasm while the second one denotes common supertype of Wasm-specific aggregate types, and the latter two are the corresponding type variants which allow nullable references. Currently, the type records look like:

\begin{lstlisting}[style=context]
v1.JSTy = object(methods:["apply",..],..)
v1.WasmTy.valTy = {externref,(ref any),..}
\end{lstlisting}

Next, \coolname may decide to introduce a Wasm object and choose the generator \texttt{WasmTableGenerator}, which constructs a \texttt{WebAssembly.Table} under JS context. This generator randomly specifies the relevant fields such as the initial table size and the type of elements as per the standard. Suppose the generator takes \texttt{externref} as element type, it then searches current generation context for variables with or matching the Wasm type \texttt{externref} to be used as table's initialization value and finds out that JS variable \texttt{v1} meets the requirement. In this way, \coolname enables JS variables to be adopted under Wasm context:

\begin{lstlisting}[style=js-small-size]
const v1 = Reflect;
const v2 = new WebAssembly.Table(
    {element:"externref", initial:4}, v1);
\end{lstlisting}

Again, the dual-type analyzer will investigate the JS and Wasm types of the new variable \texttt{v2}. The JS type of \texttt{v2} can be obtained directly from the static type environment, which is an \texttt{object(..)} type with customized methods such as \texttt{grow}. The Wasm types of \texttt{v2} are twofold. On the one hand, it is of Wasm table type with associated fields. On the other hand, it can be treated as Wasm value types the same way as \texttt{v1} does. Therefore, the type records for \texttt{v2} look like:

\begin{lstlisting}[style=context,
numbers=left,
xleftmargin=1em,
numbersep=6pt]
v2.JSTy = object(properties:["length"],..)
v2.WasmTy.tblTy = table(initial:4,elem:..)
v2.WasmTy.valTy = {externref,(ref any),..}
\end{lstlisting}

Finally, the generator scheduler might observe that the \texttt{PrototypeOverwriteGenerator} has been executed less frequently, and consequently increase its probability of being selected. This process ultimately leads to its selection by \coolname. \texttt{PrototypeOverwriteGenerator} needs two variables preferably of \texttt{object} type as inputs. As per the current type records, it picks \texttt{v1} and \texttt{v2}, resulting in JS object \texttt{v1}'s \texttt{\_\_proto\_\_} field being set to a Wasm object \texttt{v2}.

\begin{lstlisting}[style=js-small-size]
const v1 = Reflect;
const v2 = new WebAssembly.Table(
    {element:"externref", initial:4}, v1);
v1.__proto__ = v2;
\end{lstlisting}

The preceding example conceptually demonstrates how \coolname operates and crafts JS-Wasm interactions. In the following subsections, we discuss each involved component in more detail.

\subsection{Dual-Type Analysis}\label{sec:dual-type-analysis}

The sole purpose of dual-type analysis is to answer the following two questions: How can a specific Wasm type be regarded as JS types and conversely how can a specific JS type be regarded as Wasm types.

To answer these two questions, we first demand a fuzzer that understands both JS and Wasm. However, the initial version of Fuzzilli only implements JS type inference and lacks support for Wasm.  Thus, dual-type analysis first aims to incorporate Wasm type representation into the fuzzer and allow fuzzer to trace variable's Wasm types. Specifically, we begin by modeling every Wasm type involved in the cross-language interaction:

\begin{enumerate}
    \item \textbf{Value} type directly maps to its counterpart in the Wasm standard~\cite{wasm-standard}, which serves as the basis for other Wasm types. We support all spec-defined value types, namely, \texttt{i32}, \texttt{i64}, \texttt{f32}, \texttt{f64}, \texttt{v128}, and reference types.
    \item \textbf{Memory} type contains four fields that jointly describe its characteristics. The \textit{initial} and \textit{maximum} fields are used to specify the size range of storage. The \textit{addrtype} field classifies whether it is 32-bit or 64-bit addressable. Lastly, the \textit{shared} field determines if the storage can be shared across modules.
    \item \textbf{Table} type is similar to the memory type except that it additionally includes an \textit{elementType} field that specifies the reference type of contained elements.
    \item \textbf{Global} type consists of the \textit{contentType} and \textit{mutability} fields. The former refers to the value type of this global while the latter indicates if it could be modified.
    \item \textbf{Function} type encompasses the \textit{parameters} and \textit{results} fields which together define the function signature and each field is composed of the list of value types.
    \item \textbf{Tag} type is nearly identical to the function type but the \textit{results} field is kept empty as per the standard~\cite{wasm-standard}.
    \item \textbf{Module} type represents a whole Wasm module but only contains information necessary for the cross-language interaction, such as \textit{import} and \textit{export} object types, and omits irrelevant details like data segments.
    \item \textbf{Instance} type is informationally identical to the module type and is utilized solely to determine if a module has been instantiated during the generation process.
\end{enumerate}

\begin{figure}[t]
    \centering
    \includegraphics[width=\linewidth]{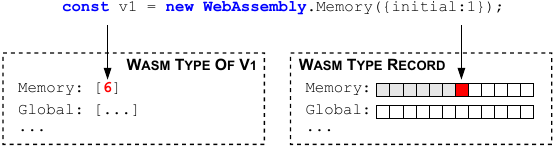}
    \caption{An illustration of Wasm type tracking in \coolname.}
    \label{fig:wasm-type-tracking}
\end{figure}

To support Wasm type tracking on top of this, we build and maintain a type record during test case generation. This record maintains a separate list for each Wasm type category based upon above classification and catalogs all encountered Wasm types. Each variable is then augmented with a Wasm type, represented as a collection of indices, in addition to its original JS type. Each index serves as a subscript to look up a specific type definition from the corresponding type list within this record. Figure~\ref{fig:wasm-type-tracking} illustrates an example where the Wasm memory type of \texttt{v1} is added to the record's \texttt{Memory} list at subscript \texttt{6}. Consequently, we add the index \texttt{6} to the \texttt{Memory} part of \texttt{v1}'s Wasm type to denote the actual Wasm memory type of \texttt{v1}. Besides, it is worth mentioning that all Wasm types must be canonicalized into unified form before being added into the type record to facilitate cross-module type matching.

\begin{table*}[t]
\centering
\caption{Wasm-to-JS Type Conversion Rules.}
\label{tab:wasm-to-js-conversion}
\renewcommand{\arraystretch}{1.15}
\resizebox{\linewidth}{!}{%
\begin{tabular}{@{}l|l@{}}
\toprule
{\textbf{Wasm Type}} & \textbf{Corresponding JS Type}                                         \\ \midrule
{\texttt{\textbf{Memory}}}                                       & \texttt{\textbf{object}(properties: ["buffer"], methods: ["grow", "toResizableBuffer", "toFixedLengthBuffer"])} \\
\texttt{\textbf{Table}}         & \texttt{\textbf{object}(properties: ["length"], methods: ["get", "grow", "set"])}        \\
\texttt{\textbf{Global}}        & \texttt{\textbf{object}(properties: ["value"], methods: ["valueOf"])}                    \\
\texttt{\textbf{Tag}}           & \texttt{\textbf{object}(properties: [], methods: [])}                                    \\
\texttt{\textbf{Module}}        & \texttt{\textbf{object}(properties: [], methods: [])}                                    \\
\texttt{\textbf{Instance}}      & \texttt{\textbf{object}(properties: ["exports"], methods: [])}                           \\ \midrule
\texttt{\textbf{Function(params, [])}} & \texttt{\textbf{function}(signature: params.map(toJSType) => undefined)}                                    \\
\texttt{\textbf{Function(params, [result])}}               & \texttt{\textbf{function}(signature: params.map(toJSType) => toJSType(result))} \\
\texttt{\textbf{Function(params, [results..])}}    & \texttt{\textbf{function}(signature: params.map(toJSType) => JSArray)}               \\ \bottomrule
\end{tabular}%
}
\end{table*}

With the support of both JS and Wasm type tracking, the dual-type analysis addresses the two questions presented at the beginning of this section by establishing type conversion rules between the two. Before digging into the details of type conversion rules, it should be noted that this type conversion process presupposes that one type should be known to infer the other. For variables originating from the JS generators, the JS type is first deduced by Fuzzilli's native type analyzer and the Wasm types are then obtained by rules. For variables originating from Wasm generators, the Wasm type comes in the first place, which is acquired by the type inference rules elaborated in Section~\ref{sec:wasm-generation}, and the JS type is later attained by conversion rules.

\begin{algorithm}[t]
    \small
    \caption{\small \textsc{Wasm Value Type Conversion Algorithm}\label{alg:wasm-val-convert}}
    \vspace{1pt}
    \begin{spacing}{1.15}
    \begin{algorithmic}[1]
        \REQUIRE{$valTy$: \textit{original Wasm value type}}
        \ENSURE{$jsTy$: \textit{corresponding JS type converted from} $valTy$}
        \PROCEDURE{\texttt{\textbf{toJSType}}}{$valTy$}
            \STATE{$jsTy \gets$ \texttt{\textbf{nothing}}}
            \SWITCH{$valTy$}
                    \CASELINE{\texttt{\textbf{i32,f32,f64}}}{ $jsTy \gets jsTy\ \cup$ \texttt{\textbf{number}}}
                    \CASELINE{\texttt{\textbf{i64}}}{ $jsTy \gets\ jsTy\ \cup$ \texttt{\textbf{bigint}}}
                    \CASELINE{\texttt{\textbf{v128}}}{ \textsc{\textbf{\texttt{unsupportedError}}}()}
                    \CASELINE{(\texttt{\textbf{ref}} $refTy$)}
                        \begin{ALC@g}
                            \IF{$refTy$ \textbf{is} \texttt{\textbf{exnref}} \textbf{or} $refTy$ \textbf{is} \texttt{\textbf{nullexnref}}}
                                \STATE{\textsc{\textbf{\texttt{unsupportedError}}}()}
                            \ENDIF
                            \SWITCH{\texttt{\textbf{\textsc{\texttt{convertToAbstractType}}}}($refTy.heapTy$)}
                                \CASELINE{\texttt{\textbf{struct,array}}}{ $jsTy \gets\ jsTy\ \cup$ \texttt{\textbf{object()}}}
                                \CASELINE{\texttt{\textbf{func}}}{$jsTy \gets\ jsTy\ \cup$ \texttt{\textbf{function()}}}
                                \CASELINE{\texttt{\textbf{i31}}}{$jsTy \gets\ jsTy\ \cup$ \texttt{\textbf{number}}}
                                \DEFAULTLINE{}{$jsTy \gets\ jsTy\ \cup$ \texttt{\textbf{anything}}}
                            \ENDSWITCH
                            \IF{$refTy.nullable$}
                                \STATE{$jsTy \gets\ jsTy\ \cup$ \texttt{\textbf{nullish}}}
                            \ENDIF
                        \end{ALC@g}
            \ENDSWITCH
            \RETURN{$jsTy$}
        \ENDPROCEDURE
    \end{algorithmic}
    \end{spacing}
\end{algorithm}

We then illustrate the type conversion rules for the first question in Table~\ref{tab:wasm-to-js-conversion} and Algorithm~\ref{alg:wasm-val-convert}, which are summarized from the standard~\cite{wasm-js-api}. As can be seen in Table~\ref{tab:wasm-to-js-conversion}, most of the Wasm types except \texttt{Function} utilize a single JS type template which is shared across all their different instances. Consequently, we incorporate these JS type templates into the static type environment which allows fast JS type lookup for these Wasm types. Besides, while omitted from Table~\ref{tab:wasm-to-js-conversion}, we also provide type annotations for properties and methods. For instance, the \texttt{buffer} property within the JS type for \texttt{Memory} is tagged as \texttt{jsArrayBuffer}. \texttt{Function} types present a more complex case. Their signatures vary and are defined by Wasm value types. Thus, deriving corresponding JS function signature demands converting these Wasm types into their respective JS types. Algorithm~\ref{alg:wasm-val-convert}  details the case-by-case conversion process for each input Wasm value type $valTy$. Most type conversions presented in Algorithm~\ref{alg:wasm-val-convert} are intuitive, with the exception of reference types, which need further explanation. Specifically, the JS type of a reference type $refTy$ depends on the contained heap type. There are two kinds of heap types in the current Wasm standard~\cite{wasm-standard}. Concrete heap types refer to specific \texttt{struct}, \texttt{array}, and \texttt{func} type definitions declared in the module while abstract heap types serve as general supertypes representing classes of related types. To enable a unified conversion process, the \texttt{convertToAbstractType} function first maps concrete types to their corresponding abstract supertypes. Algorithm ~\ref{alg:wasm-val-convert} subsequently operates exclusively on these abstract types to determine the final JS type. For example, a reference type that refers to a concrete \texttt{struct} type definition is handled in the same way as the type that points to abstract \texttt{struct} type, and thereby converted to an exotic JS object type with no properties and methods. Besides, for nullable reference type, we augment its corresponding JS type with \texttt{nullish} to account for potential null values.

On the other hand, the type conversion from JS to Wasm follows a similar procedure. The primary caveat, however, is that JS types are mostly restricted to mapping only to the Wasm value types. Algorithm~\ref{alg:js-convert} lists the simplified JS type conversion process for addressing the second question and is largely self-explanatory. For instance, Algorithm~\ref{alg:js-convert} allows a JS \texttt{integer} type to be interpreted as a Wasm reference type pointing to \texttt{i31} heap type, which denotes an unboxed scalar. Moreover, it is important to note that a Wasm object type can be further converted to a Wasm value type here via its previously inferred JS type, thereby broadening the set of available Wasm types.

At last, it is worth mentioning that \texttt{v128}, \texttt{exnref}, and \texttt{nullexnref} types are purposefully excluded from type conversion process, since they are explicitly unsupported by the interaction specification~\cite{wasm-js-api}. Accordingly, we also avoid these types when generating Wasm elements, which will be further detailed in Section~\ref{sec:wasm-generation}.

\begin{algorithm}[bt]
    \small
    \caption{\small \textsc{JS Type Conversion Algorithm}\label{alg:js-convert}}
    \vspace{1pt}
    \begin{spacing}{1.15}
    \begin{algorithmic}[1]
        \REQUIRE{$jsTy$: \textit{original JS type}}
        \ENSURE{$valTy$: \textit{set of Wasm value types representing $jsTy$}}
        \PROCEDURE{\texttt{\textbf{toWasmValueTypes}}}{$jsTy$}
            \STATE{$valTy \gets $ \{\}}
            \IF{$jsTy$ \textbf{is} \texttt{\textbf{nullish}}}
                \FOR{$absHeapTy \in absHeapTypes$}
                    \STATE{$valTy \gets valTy\ \cup$ \texttt{\textbf{(ref null $absHeapTy$)}}}
                \ENDFOR
                \STATE{$\ .\ .\ .$}
            \ENDIF
            \IF{$jsTy$ \textbf{is} \texttt{\textbf{integer}}}
                \STATE{$valTy \gets valTy\ \cup$ \{\textbf{\texttt{i32,f32,f64,(ref i31)}}\}}
                \STATE{$valTy \gets valTy\ \cup$ \textbf{\texttt{(ref null i31)}}}
            \ENDIF
            \IF{$jsTy$ \textbf{may not be} \textbf{\texttt{nullish}}}
                \STATE{$valTy \gets valTy\ \cup$ \{\textbf{\texttt{(ref any),(ref extern)}}\}}
                \STATE{$valTy \gets valTy\ \cup$ \textbf{\texttt{(ref null any)}}}
                \STATE{$valTy \gets valTy\ \cup$ \textbf{\texttt{(ref null extern)}}}
            \ENDIF
            \STATE{$\ .\ .\ .$}
            \STATE{$valTy \gets valTy\ \backslash$ \{\textbf{\texttt{v128,exnref,nullexnref}}\}}
            \RETURN{$valTy$}
        \ENDPROCEDURE
    \end{algorithmic}
    \end{spacing}
\end{algorithm}

\subsection{Wasm-Specific Generation}\label{sec:wasm-generation}

In this section, we provide details about the generation of Wasm-specific elements and the inference of their Wasm types which are left unspecified in Section~\ref{sec:dual-type-analysis}. Specifically, as mentioned in Section~\ref{sec:background-interaction}, certain Wasm elements can be directly constructed via the JS APIs. To accommodate this, we first design the following generators, which operates by directly invoking APIs such as \texttt{WebAssembly.Memory} to craft corresponding Wasm objects:

\begin{itemize}
    \item \texttt{WasmMemoryGenerator} generates a random Wasm memory object. It first determines whether the memory is shared and 64-bit addressable and then chooses an initial size between zero and a maximum allowed limit. Lastly, it may optionally assign a maximum size for this memory whose value is bounded by the initial size and the maximum allowed limit.
    \item \texttt{WasmTableGenerator} functions in a similar way to the \texttt{WasmMemoryGenerator}, except that Wasm table object additionally requires an element type and an initialization value. Since the relevant API currently limits the element type to \texttt{funcref} and \texttt{externref}, the generator randomly selects one of the two and then searches the generation context for a variable matching the required type. If a suitable variable exists, it is used as the initialization value. Otherwise, the type's default value is used.
    \item \texttt{WasmGlobalGenerator} generates a random Wasm global object. The value type of this global is selected arbitrarily from the list of allowed value types, namely, \texttt{i32}, \texttt{i64}, \texttt{f32}, \texttt{f64}, \texttt{funcref}, and \texttt{externref} and the mutability of this global is randomly assigned. Additionally, this generator adopts an approach similar to \texttt{WasmTableGenerator} to provide Wasm global with an initial value.
    \item \texttt{WasmTagGenerator} assembles a random Wasm tag object by providing a list of Wasm value types as tag's parameter type. Similarly, each value type is also drawn randomly from the aforesaid list of allowed value types.
\end{itemize}

Besides, given that Wasm instances serve as a primary medium for JS-Wasm interaction and that Wasm functions must be exported through these instances, we thereby design the \texttt{WasmInstanceAndExportGenerator} which aims to build Wasm instances and export contained Wasm objects. Specifically, the generator operates in three main stages: \textcircled{1} It produces an instantiable Wasm module that specifies the object types to be imported and exported. Since \coolname's goal is to generate interactions rather than fully functional Wasm module, we use a delegation-based design where the generator constructs only the module shape, which contains essential information like import and export specifications, while delegating the task of generating a concrete and shape-conforming Wasm module to third-party tools. For instance, we may only declare the function signature of an exported Wasm function but leaves function body implementation to the external tool. This design enables \coolname to leverage different Wasm module generation backends. Building upon this foundation, \coolname only needs to tackle the module shape generation problem which involves four steps. First, concrete type definitions are randomly generated, which can serve as concrete heap types for subsequent reference type generation. Second, import object types are generated. The generator randomly selects one of the five importable Wasm object kinds and yields a corresponding object type. Notably, to strengthen Wasm module's connection to the surrounding environment, the generator queries the generation context for available Wasm object types held by existing variables and randomly selects one as the import object type. Moreover, it is worth mentioning that the interaction specification~\cite{wasm-js-api} allows any variable with the Wasm value type \texttt{X} to be treated as an immutable global of the same value type \texttt{X} and any JS function to be cast to a Wasm function with arbitrarily assigned signature. This mechanism offers greater flexibility for the generator when generating import types for global and function. Third, the generator randomly produces Wasm object types to act as candidates for subsequent export. This generation process mirrors the methodology employed by the previous generators such as \texttt{WasmGlobalGenerator}, with the primary distinction being the unrestricted selection of Wasm value types. Fourth, the generator selects a subset of the currently available Wasm object types to be exported. During this selection, any Wasm object types involving the \texttt{v128}, \texttt{exnref}, or \texttt{nullexnref} value types are excluded from consideration, as previously mentioned in Section~\ref{sec:dual-type-analysis}. \textcircled{2} The generator then instantiates the newly produced Wasm module. To satisfy each required import, it randomly picks a type-compatible variable from the generation context. \textcircled{3} As the final step, the generator exports the Wasm instance's complete set of exports to surrounding environment, making them available to subsequent code generators.

This concludes the description of \coolname's primary Wasm generators. As evident from the foregoing discussion, the Wasm type inference for the generated variables should be trivial to implement, since all requisite Wasm type information has been inherently captured during the generation process itself. For instance, the Wasm value type generated in the \texttt{WasmGlobalGenerator} directly corresponds to the \textit{contentType} field of \texttt{Global} type. With the support of Wasm generators and dual-type analysis, \coolname can now seamlessly generate cross-language interactions. The current challenge lies in ensuring interaction diversity, which will be addressed in Section~\ref{sec:scheduling}.

\subsection{Performance-Based Scheduler}\label{sec:scheduling}

The original version of Fuzzilli allocates fixed and static selection weights to all involved generators. This approach, while probably suitable for pure JS code generation, proves ill-suited for the JS-Wasm interaction context. The core issue originates from the significant disparity in the number of JS versus Wasm generators. If Wasm generators are assigned weights comparable to their numerous JS counterparts, their selection probability becomes exceedingly low. Conversely, assigning them disproportionately high weights introduces excessive runtime overhead and frequent timeouts, which is observed in Section~\ref{sec:ablation}. Moreover, static allocation scheme hinders the exploration of infrequently-selected generators during the fuzzing campaign. Therefore, a dynamic weight adjustment strategy is demanded.

To solve this, we adopt an established approach from prior work~\cite{slime,ecofuzz}, framing the fuzzing process as a multi-armed bandit problem where each arm denotes a generator. During the fuzzing loop, we adjust the selection weights to prioritize generators that have returned higher performance rewards previously and those that remain insufficiently explored. To implement this, we first define what constitutes the generator performance. An initial, immature metric may be the proportion of path-discovering test cases. However, our preliminary experiment results indicate that the fuzzer at most of the time generates test cases that fail to trigger new code paths. This renders such a metric ineffective, as it lacks sufficient discriminative power. Therefore, we employ a relatively lenient metric: the proportion of valid test cases contributed by a generator, which is defined as:

\begin{equation}
    perf[i]=\frac{Num_{valid}[i]}{Num_{total}[i]}
\end{equation}

The $Num_{valid}[i]$ denotes the number of valid test cases generated by the $i$-th generator, which also contains path-discovering test cases. The $Num_{total}[i]$ represents the total number of test cases generated by the $i$-th generator, which additionally includes failed, timeout, and crash test cases. This metric serves a dual purpose: it incentivizes the generation of valid test cases which are more likely to trigger deep code logic, which simultaneously penalizing generators that produce excessive timeouts.

We then use the UCB-1 algorithm to dynamically adjust the selection weights of the generators based on this defined performance metric and the historical selection frequency:

\begin{equation}
    Num_{total} =\sum_{i=0}^{n-1}Num_{total}[i]
\end{equation}

\begin{equation}
    weight[i]=(perf[i]+\sqrt{\frac{2.0 *ln(Num_{total})}{Num_{total}[i]}})*100
\end{equation}

Specifically, the $Num_{total}$ denotes the total number of test cases generated by all $n$ generators and the output of the UCB-1 algorithm is multiplied by 100 before serving as the final result to align with the order of magnitude of the original weights used in Fuzzilli. Moreover, the computed weights for the Wasm generators are amplified by a constant factor to guarantee a sufficient supply of Wasm objects for interaction generation within a single test case. Finally, we also incorporate the above process into the mutators, which enables \coolname to intelligently schedule generators and mutators based on the progress of fuzzing campaign.
\section{Implementation}\label{sec:implementation}

\begin{figure}[htbp]
    \centering
    \includegraphics[width=\linewidth]{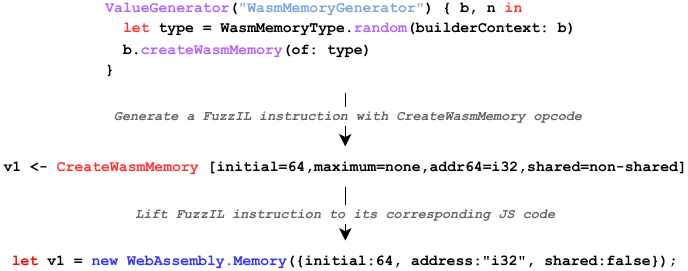}
    \caption{The general workflow of Wasm element generation in \coolname.}
    \label{fig:fuzzil-implementation}
\end{figure}

We have developed the prototype of \coolname on top of Fuzzilli~\cite{Fuzzilli} (commit \texttt{0e20cd55}) with more than 8K lines of Swift code added. To support the straight generation of Wasm-related elements, we extend FuzzIL, the intermediate representation used by Fuzzilli, with new opcodes targeting the Wasm components and implement the serialization, type tracking, and lifting support for these opcodes. This allows \coolname to handle these newly introduced Wasm variables in the same way as native JS variables. A series of generators are then designed to produce opcode-corresponding FuzzIL instructions. Figure~\ref{fig:fuzzil-implementation} shows an example where the generator \texttt{WasmMemoryGenerator} introduces a Wasm memory object into the test case that is currently under construction, which is represented as a FuzzIL instruction with the opcode \texttt{CreateWasmMemory}. When the generation process ends, the FuzzIL instruction is lifted to its corresponding JS code snippet for execution. Besides, we leverage wasm-smith~\cite{wasm-smith} (commit \texttt{62cdf1ce}) to help \coolname generate valid Wasm modules with the specified imports and exports. Specifically, the \texttt{module\_shape} option of wasm-smith accepts a Wasm binary module and enforces the generated Wasm module to have the same imports and exports as those in the passed-in module but allows different implementation for other parts. Since wasm-smith is written in Rust, we use its C bindings and incorporate it into \coolname as a shared library. In order to communicate with wasm-smith and interpret the returned Wasm module, \coolname reuses the parser and encoder from WasmKit~\cite{wasmkit} to parse Wasm binary module into the internal representation and vice versa. However, WasmKit currently lacks support for more recent Wasm standards such as multi-memory or garbage collection. Therefore, we implement the support for these standards manually in \coolname.

In addition to the above discussion about Wasm element generation, the dual-type analyzer is implemented on top of Fuzzilli's JS typer and static type environment is extended from Fuzzilli's JS static environment, which combinatorially allow \coolname to track the JS and Wasm types of each new variable in a flow-sensitive manner. Moreover, the scheduler is registered as periodical task at fuzzer's initialization stage which triggers every 60 minutes and recalculates the weights for mutators and generators.

Lastly, there are a few additional remarks regarding the implementation of \coolname. Firstly, \coolname enables all Wasm 3.0 features by default except JS String Builtins~\cite{js-string-builtin}, which demands further handling. It is worth noting that not all Wasm features are well supported across all JS engines or hardware architectures. For instance, JavaScriptCore does not support memory64 standard currently. More discussion on feature support is elucidated in the evaluation. Secondly, the implementation actually includes specific modifications to existing mutators, though these are not explicitly claimed as a contribution. For example, the splice mutator is used to incorporate a program slice from other program into the current program. To reinforce data flow correlation between the two, splice mutator remaps a subset of slice's variables to variables of matching types in current program. However, the original implementation only considers the JS types of variables when performing type matching, thereby raising a lot of runtime errors. \coolname adjusts the implementation to take both the JS types and Wasm types into account when remapping variables.

\section{Evaluation}\label{sec:evaluation}

To validate the effectiveness of our design, we evaluate \coolname and answer the following four questions:
\begin{enumerate}[label=\textbf{RQ\arabic*:}, ref={RQ\arabic*}, leftmargin=0ex, labelsep=\widthof{~}, itemindent=\widthof{RQ2:~\hspace{.125ex}}]
    \item \label{RQ1} How does \coolname perform in terms of code coverage and bug findings when compared to the state-of-the-art fuzzers? (see Section~\ref{sec:comparison})
    \item \label{RQ2} Can \coolname generate correct interaction code, both syntactically and semantically? (see Section~\ref{sec:validity})
    \item \label{RQ3} What is the contribution of each individual component to the performance of \coolname? (see Section~\ref{sec:ablation})
    \item \label{RQ4} Can \coolname find new bugs in latest JS engines? (see Section~\ref{sec:real-bug})
\end{enumerate}

\subsection{Experimental Setup}\label{sec:evaluation-setup}

We conduct all the experiments on one machine running Ubuntu 20.04 x86-64 with Intel Xeon Platinum 8380 CPUs (2.3GHz, 160 cores) and 256GB main memory. To facilitate the isolation of execution environments and the management of hardware resources, all experiments are further executed in Docker containers.

\noindent\textbf{Baselines.} Given the scarcity of fuzzing works targeting the JS-Wasm interactions, we choose only two fuzzing methods as baselines to evaluate the efficacy of \coolname: \textcircled{1} Fuzzilli ~\cite{Fuzzilli} (commit \texttt{0e20cd55}) represents the state-of-the-art JS fuzzer which also serves as the basis for \coolname. Besides, we have noticed that when developing \coolname, Fuzzilli's maintainers are simultaneously integrating Wasm support to Fuzzilli. Therefore, we include one newer version of Fuzzilli (commit \texttt{675eccd6}) into comparison, which can construct cross-language interactions similar to \coolname. \textcircled{2} Dharma ~\cite{dharmafuzz} is a generative fuzzer developed by Mozilla which uses context-free grammars to generate test inputs. FuzzingLabs has previously contributed a grammar file~\cite{dharma-grammar} that enables Dharma to explore basic Wasm-JS APIs~\cite{wasm-js-api}. Consequently, we select Dharma, along with this specific grammar file, as the baseline for comparison.

\noindent\textbf{Benchmarks.} We choose three mainstream JS engines that come with Wasm support: Firefox SpiderMonkey~\cite{spidermonkey} (commit \texttt{55279d59}), Chrome V8~\cite{v8} (commit \texttt{118ebe01}), and Safari JavaScriptCore~\cite{javascriptcore} (commit \texttt{736f1346}). It is worth noting that all these engines have been heavily tested by the security researchers and continuous fuzzing projects such as OSS-Fuzz~\cite{oss-fuzz} and we use the latest commit available at the time of our evaluation for each engine.

\noindent\textbf{Configurations.} Given that all involved fuzzers support test case generation without seeds, we do not provide initial seed corpus for each fuzzer. This enables a fair comparison since Dharma is a pure generation-based approach. Moreover, we build all JS engines with default configurations provided by Fuzzilli. Similarly, we leverage the command-line arguments Fuzzilli supplies to each JS engine, which alter settings such as JIT trigger threshold. Lastly, we set the timeout of a single execution to 600 ms to address potential infinite loops and repeat each experiment 5 times to mitigate the randomness. More details on configurations will be discussed in relevant subsections.

\subsection{Comparison to State-of-the-Art Fuzzers}\label{sec:comparison}

To address~\ref{RQ1}, we have run \coolname and other state-of-the-art fuzzers on three JS engines as specified in Section ~\ref{sec:evaluation-setup} over a period of 72 hours. This aligns with test durations of previous works~\cite{OptFuzz,JITPicking}. Given that JavaScriptCore lacks support for memory64, multiple-memory, and relaxed-SIMD standards currently, relevant Wasm instructions and settings are excluded when testing JavaScriptCore. Moreover, unlike \coolname and Fuzzilli which are armed with native coverage collection support from three JS engines and advanced crash detection, Dharma is merely a test case generator that yields the specified number of test cases. Considering the limited disk space, it is impractical to let Dharma run continuously for three days and save all generated test cases for coverage calculation and bug detection. Therefore, we have written a custom script in Python which performs the following steps: \textcircled{1} The script instructs Dharma to generate a number of test cases under the specified directory; \textcircled{2} When the generation process finishes, the script runs each test case sequentially on the instrumented JS engine to collect raw coverage data. Besides, the script monitors the exit code of each execution and saves any test cases raising non-zero exit codes for later analysis; \textcircled{3} The raw coverage data is merged with previous coverage profile to produce the new profile which contains accumulated coverage. \textcircled{4} Once the merge process is done, the script deletes all generated test cases and enters the next round until the time budget is reached.

To enable the accumulative coverage calculation, we use the llvm-cov toolchain~\cite{llvm-cov}. Specifically, we instrument the JS engine with source-based code coverage. This enables JS engine to emit a \texttt{profraw} file which encodes raw coverage information during execution. The \texttt{profraw} files can then be merged by llvm-profdata into one \texttt{profdata} file which contains accumulated coverage. To ensure the comparability of coverage results, we apply the same coverage calculation methodology to \coolname and Fuzzilli, namely, we sort the finally generated seed corpus chronologically and calculate the accumulative coverage.

\begin{table*}[t]
\centering
\caption{The coverage results averaged across five runs, as measured by llvm-cov. Each field consists of four parts, representing region, function, line, and branch coverage. The best result is marked in bold.}
\renewcommand{\arraystretch}{1.15}
\label{tab:coverage}
\resizebox{\linewidth}{!}{%
\begin{tabular}{l | l @{ / } l @{ / } l @{ / } ll @{ / } l @{ / } l @{ / } ll @{ / } l @{ / } l @{ / } l}
\toprule
\textbf{Fuzzer} & \multicolumn{4}{c}{\textbf{JavaScriptCore}} & \multicolumn{4}{c}{\textbf{V8}} & \multicolumn{4}{c}{\textbf{SpiderMonkey}} \\
\midrule

\coolname
  & \bfseries 35.05\% & \bfseries 46.73\% & \bfseries 34.67\% & \bfseries 37.92\%
  & \bfseries 30.34\% & \bfseries 42.40\% & \bfseries 35.10\% & \bfseries 32.20\%
  & \bfseries 34.58\% & \bfseries 55.91\% & \bfseries 44.94\% & \bfseries 36.46\% \\

Fuzzilli-\texttt{0e20cd55}
  & 32.76\% & 45.55\% & 33.03\% & 35.94\%
  & 29.26\% & 40.73\% & 33.89\% & 30.87\%
  & 31.67\% & 51.39\% & 41.40\% & 33.33\% \\

Fuzzilli-\texttt{675eccd6}
  & 34.47\% & 45.66\% & 34.46\% & 37.72\%
  & 30.14\% & 42.04\% & 35.07\% & 31.95\%
  & 34.30\% & 55.35\% & 44.60\% & 36.25\% \\

Dharma
  & 09.92\% & 17.61\% & 08.76\% & 06.63\%
  & 11.73\% & 18.64\% & 12.34\% & 10.38\%
  & 11.83\% & 20.61\% & 13.54\% & 10.04\% \\
\midrule

\coolname-no-scheduler
  & 34.01\% & 46.08\% & 33.42\% & 36.39\%
  & 30.32\% & \bfseries 42.51\% & 34.97\% & 32.18\%
  & 33.03\% & 53.69\% & 42.59\% & 34.49\% \\
\bottomrule
\end{tabular}
}
\end{table*}

\begin{figure*}[htbp]
    \centering
    \includegraphics[width=\linewidth]{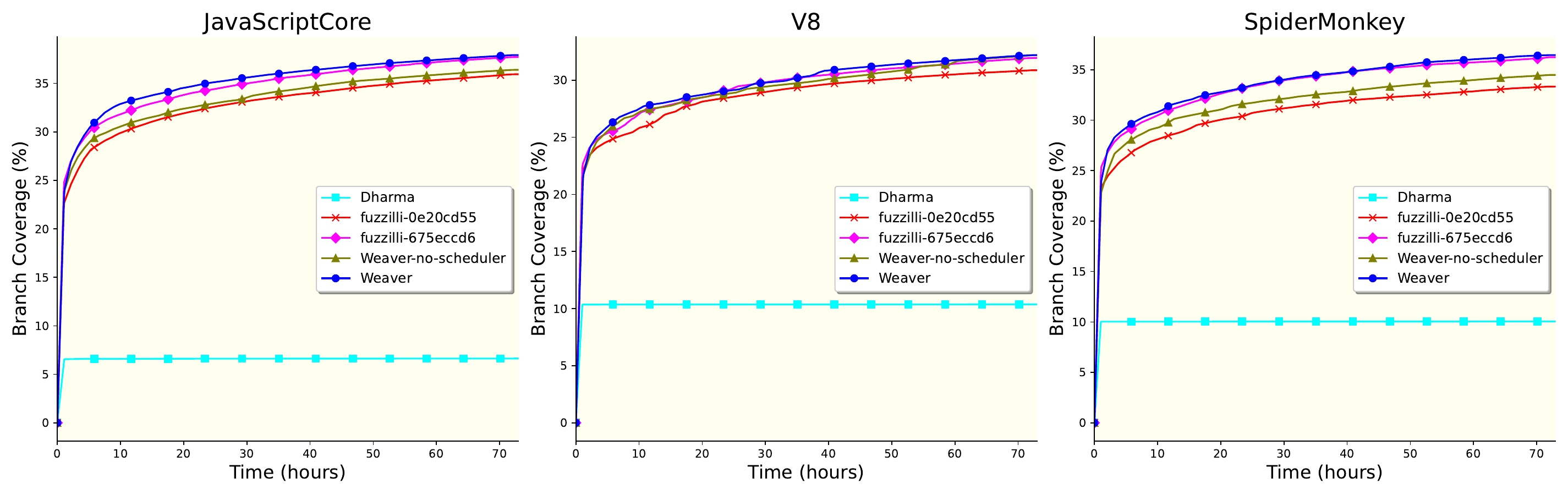}
    \caption{The branch coverage for all the evaluated techniques over time.}
    \label{fig:coverage}
\end{figure*}

Table~\ref{tab:coverage} demonstrates the average percentage of region, function, line, and branch coverage achieved across three JS engines by relevant fuzzers and Figure~\ref{fig:coverage} shows the branch coverage evolution over time. For all JS engines, \coolname consistently outperforms other competing fuzzers in all four coverage metrics. When compared to the earlier version of Fuzzilli on which \coolname is built, \coolname finds at most 9.2\% more regions, 8.8\% more functions, 8.6\% more lines, and 9.4\% more branches, a result primarily attributed to the introduction of JS-Wasm interaction in \coolname. While the newer version of Fuzzilli has incorporated the generation of Wasm elements and devised several new generators such as \texttt{UnboundFunctionCallGenerator}, its achieved code coverage remains slightly inferior to that of \coolname. And it is worth mentioning that different from the methods adopted by Fuzzilli which directly generates Wasm instructions at the intermediate representation level, \coolname uses a decoupled design where the generation of Wasm modules is delegated to third-party tools. This brings two advantages: First, this grants \coolname the extra flexibility to utilize Wasm module generators other than wasm-smith, benefiting \coolname from different generation algorithms. Second, this design obviates the needs for \coolname to continually update its opcodes to align with the latest proposals given the evolving nature of Wasm. In particular, these frequent updates have caused the newer version of Fuzzilli to crash during our evaluation. At last, Dharma's coverage performance across all JS engines is suboptimal, and it fails to exhibit an increasing trend over time. This is expected since Dharma's effectiveness heavily depends on the provided grammar and the current grammar designates only one fixed Wasm module and wraps all Wasm element constructions into \texttt{try-catch} statements, which might alter the semantics of code snippets and suppress JIT optimizations according to CodeAlchemist~\cite{CodeAlchemist}.

Unfortunately, none of the evaluated fuzzers are able to discover new bugs in the JS engines within the 3-day period. We deem this as acceptable because all involved JS engines have been continuously fuzzed by their internal fuzzers and it is unlikely to find new crashes in the currently set fuzzing period. This hypothesis is further validated by the results of prior works~\cite{OptFuzz, Waltzz}. For instance, a two-week bug finding comparison experiment conducted in OptFuzz~\cite{OptFuzz} is unable to uncover any bugs. Consequently, we resort to a long-term bug finding campaign to prove the efficacy of \coolname. The details are discussed in Section~\ref{sec:real-bug}.

\subsection{Validity of Generated Inputs}\label{sec:validity}

\begin{table}[t]
\centering
\caption{The average validity and timeout rates of relevant fuzzers. The best result is marked in bold.}
\label{tab:validity-timeout}
\renewcommand{\arraystretch}{1.15}
\resizebox{\columnwidth}{!}{%
\begin{tabular}{@{}lr|ccc@{}}
\toprule
\multicolumn{2}{l|}{\textbf{JS Engine}} & \textbf{\coolname} & \textbf{Fuzzilli-\texttt{0e20cd55}} & \textbf{\coolname-no-scheduler} \\ \midrule
\multirow{2}{*}{JavaScriptCore} & \textit{Validity} & 63.52\% & \textbf{71.17\%} & 58.54\% \\
                                & \textit{Timeout}  & 09.55\% & \textbf{03.86\%} & 16.27\% \\ \midrule
\multirow{2}{*}{V8}             & \textit{Validity} & 67.51\% & \textbf{71.92\%} & 66.71\% \\
                                & \textit{Timeout}  & 12.65\% & \textbf{04.71\%} & 25.25\% \\ \midrule
\multirow{2}{*}{SpiderMonkey}   & \textit{Validity} & 65.91\% & \textbf{70.45\%} & 67.19\% \\
                                & \textit{Timeout}  & 06.08\% & \textbf{04.08\%} & 11.02\% \\ \bottomrule
\end{tabular}%
}
\end{table}

The validity of test cases has long been a core concern for JS fuzzers, as it directly impacts the triggering of deeper code paths such as those related to JIT optimization. In this section, We consider a test case as valid if its execution on JS engine does not throw any runtime errors.

To answer \ref{RQ2}, we record the number of valid test cases along with the number of non-timeout test cases and define the validity rate as the division of the two. Table~\ref{tab:validity-timeout} lists the validity rates of related fuzzers averaged across five 72-hour runs. First, we observe that even Fuzzilli, the state-of-the-art JS fuzzer, cannot guarantee the validity of generated inputs. This is mainly attributed to dynamic and complex nature of JS language and the static type analysis employed in Fuzzilli may fail to correctly infer the variable types. The integration of Wasm further worsens this situation since it adds an extra layer of validity checks. This is confirmed by the results in Table~\ref{tab:validity-timeout} where the validity rates of \coolname are lower than that of Fuzzilli. Nevertheless, the average decrease is under 6\% owning to the dual-type analysis design, which we deem acceptable. We enumerate several typical causes of runtime errors as follows:

\noindent\textcircled{1} \textbf{The limitation of wasm-smith.} Wasm-smith ensures the generated modules to pass static validation, but provides no guarantee regarding their runtime behaviors. Therefore, we occasionally witness \texttt{RuntimeError} raised when running exported Wasm functions or instantiating Wasm modules. As an example, the following module built by wasm-smith will report an out-of-bound table index error during instantiation. Wasm allows for the active initialization of specified ranges of a table using element segments at instantiation time. The problem here is that wasm-smith uses the value of a global variable, which is set to a negative number, as the offset for table initialization, thereby causing a runtime error. Besides, it is worth mentioning that it is infeasible for Wasm module generators to statically determine if a module will throw an error. Thus, we consider this kind of error as unavoidable.

\begin{lstlisting}[style=js-small-size]
// The truncated Wasm module:
// (module
//   (table (;0;) 1937 externref)
//   (global (;2;) i32 i32.const -1051341)
//   (elem (table 0) (global.get 2) ..)
// )
const v1 = new WebAssembly.Module(..);
// RuntimeError:
// table index is out of bounds.
const v2 = new WebAssembly.Instance(v1);
\end{lstlisting}

\noindent\textcircled{2} \textbf{The drawback of static type environment.} Currently, \coolname encodes the JS types of most Wasm elements into the static type environment to enable fast type lookup, which aligns with Fuzzilli's treatments for built-ins. However, the types of some object properties may vary depending on the arguments provided during object construction. To cover all possible type cases, we generally assign a more general type to these properties statically, which results in type mismatch at runtime. For example, we can access the underlying value of \texttt{WebAssembly.Global} object using \texttt{value} property and the type of this property can be \texttt{number}, \texttt{BigInt}, or \texttt{object} depending on the constructor arguments. To cover these possibilities, we designate the \texttt{anything} type to this property which represents any possible types. As some code generators may select a potentially compatible variable from the context when a variable of required type is unavailable, this could lead to \texttt{TypeError} if the generator chooses this property.

\begin{lstlisting}[style=js-small-size]
const v1 =
    new WebAssembly.Global({value:"i64"});
// TypeError:
// Cannot convert a BigInt to a number.
Date.UTC(v1.value);
\end{lstlisting}

\noindent\textcircled{3} \textbf{Lack of value tracking.} As discussed in Section~\ref{sec:dual-type-analysis}, a JS variable of \texttt{integer} type can be viewed as an unboxed scalar under Wasm, namely, in the form of \texttt{(ref i31)} or \texttt{i31ref}. However, the allowable value range for these two Wasm types is limited to 31 bits. Any values exceeding this range could raise a \texttt{LinkError} or \texttt{TypeError}. Ensuring the selected variable's value is within the permissible range would require value tracking, which \coolname currently does not support due to performance considerations. This makes \coolname susceptible to above exceptions. Still, we consider this limitation acceptable as such cases are rare.

\noindent\textcircled{4} \textbf{Imprecision in the type system.} In current design, any imprecision in the JS type directly propagates to the inferred Wasm types. One notable example has been observed which lies in Fuzzilli's handling of \texttt{undefined} and \texttt{null} values. Fuzzilli treats these two values as the same \texttt{nullish} type which brings up confusion in \coolname. Specifically, \texttt{null} value can be typed as any nullable reference types in Wasm while \texttt{undefined} value can not. However, since they are annotated with the same type, current type-aware generation is not able to distinguish one from the other, leading to the misuse of \texttt{undefined} and triggering of \texttt{TypeError}.

\noindent\textcircled{5} \textbf{Side effects from other generators or mutators.} Some generators might change the semantics of built-ins, compromising the correctness of subsequently generated code. For instance, \texttt{BuiltinOverwriteGenerator} can reassign the Wasm built-in \texttt{WebAssembly} to a float value, causing the next Wasm element generation to raise a \texttt{TypeError}. Similarly, mutators like operation mutator may replace the original built-in with a randomly selected one, invalidating subsequent code that depends on the original built-in.

\begin{lstlisting}[style=js-small-size]
// Executing BuiltinOverwriteGenerator.
WebAssembly = 926.7333937898886;
// TypeError:
// WebAssembly.Memory is not a constructor.
new WebAssembly.Memory({initial:63});
\end{lstlisting}

\subsection{Ablation Study}\label{sec:ablation}

To tackle~\ref{RQ3}, we partition our overall design into two main components based on the challenges identified in this paper. The first fragment consists of dual-type analysis and the JS-Wasm interaction generation informed by it and the second fragment contains the performance-based scheduler. For the ablation study, we create \coolname-no-scheduler, an ablated version of \coolname that removes the scheduler. By sequentially comparing Fuzzilli-\texttt{0e20cd55}, \coolname-no-scheduler, and the \coolname, we can assess the contribution of each component.

We evaluate \coolname-no-scheduler with the same setup as in Section~\ref{sec:comparison}. Table~\ref{tab:coverage} and Figure~\ref{fig:coverage} present its coverage results, which indicate that both components contribute to the coverage growth. Interestingly, our initial expectation is that the JS-Wasm interaction should have a greater impact on code coverage than the scheduler, given that it introduces new execution logic. However, this assumption does not hold under certain cases according to Table~\ref{tab:coverage}. For instance, the scheduler produces a 1.53\% increase in branch coverage on JavaScriptCore, compared to the mere 0.45\% increase from JS-Wasm interaction.

To seek a potential explanation for this phenomenon, we further investigate the implications of the scheduler on test case validity and timeout rates. The results are summarized in Table~\ref{tab:validity-timeout}. Surprisingly, a significant increase in the timeout rate is observed when the scheduler is removed, with about one-quarter of samples timing out in the worse case. Upon a closer inspection, we find out that most timeout samples involve constructing multiple \texttt{WebAssembly.Memory} or \texttt{WebAssembly.Table} objects with large initial sizes. In particular, creating large \texttt{WebAssembly.Table} objects in test cases incurs substantial overhead for JS engines, which can easily result in timeout. When the scheduler is disabled, the fuzzer selects generators based on the statically assigned weights. Considering the significant disparity between the number of JS and Wasm generators, we assign higher static weights for Wasm generators to balance the selection probabilities of two languages. This results in the prolific creation of objects like \texttt{WebAssembly.Table}, ultimately causing frequent timeouts. The introduction of scheduler effectively mitigates the timeout issue. This is because as the generator produces more timeout samples, its correctness rate declines, leading to a lower weight being assigned by scheduler. In addition to helping mitigate timeout rate, the scheduler also benefits the validity rate. Apart from a modest decrease on SpiderMonkey, we observe a general enhancement in the validity rate when scheduler is enabled. This result aligns with the primary objective of the scheduler. In summary, we believe that the scheduler is instrumental in the above two aspects, enabling \coolname to explore more code paths on the basis of JS-Wasm interaction generation.

\subsection{Real-World Bugs Found}\label{sec:real-bug}

 To answer~\ref{RQ4}, we conduct long-term bug discovery on three JS engines. More specifically, we leverage Fuzzilli's built-in support for distributed fuzzing and run 10 \coolname instances simultaneously in the localhost network where one fuzzer instance serves as the root node which shares seed corpus with children nodes and the other nine instances act as leaf nodes which connect to the root node. This enables a more efficient fuzzing practice. Moreover, we notice that V8 and SpiderMonkey have implemented simulators for some architectures. Therefore, we additionally build and test these two engines across different architectures. However, support for certain Wasm standards is architecture-dependent and we thereby adapt \coolname for different architectures. Table~\ref{tab:architecture-support} lists the involved architectures and their unsupported Wasm standards.

\begin{table}[t]
\centering
\caption{Overview of target JS engines, architectures, and Wasm standard support in long-term bug discovery.}
\label{tab:architecture-support}
\renewcommand{\arraystretch}{1.15}
\resizebox{\columnwidth}{!}{%
\begin{tabular}{@{}ll|l@{}}
\toprule
\textbf{JS Engine}                     & \textbf{Architecture}           & \textbf{Unsupported Wasm Standard}                  \\ \midrule
JavaScriptCore                & x86-64                 & Multiple-Memory, Memory64, Relaxed-SIMD \\ \midrule
V8                            & x86-64, riscv64, arm64 & None                                    \\ \midrule
\multirow{2}{*}{SpiderMonkey} & x86-64                 & None                                    \\
                              & riscv64, loong64       & SIMD, Relaxed-SIMD                      \\ \bottomrule
\end{tabular}%
}
\end{table}

Over approximately one month of testing, \coolname has found two previously unknown bugs and several crashes that have been patched in recent revisions. Table~\ref{tab:real-world-bugs} presents the relevant bug information. Notably, the second bug is initially discovered in the riscv64 version of SpiderMonkey but the subsequent analysis reveals that it also influences the arm64 architecture and allows random code execution. Thus, this bug is deemed a critical security vulnerability and assigned the highest priority for remediation. This demonstrates the practicality of \coolname in finding real-world bugs.

\begin{table*}[t]
\centering
\caption{Real-world bugs found by \coolname. * means this bug is security-sensitive.}
\label{tab:real-world-bugs}
\renewcommand{\arraystretch}{1.15}
\resizebox{\linewidth}{!}{%
\begin{tabular}{@{}llllllll@{}}
\toprule
\textbf{JS Engine}      & \textbf{ID} & \textbf{Issue}    & \textbf{Architecture} & \textbf{Status} & \textbf{Priority} & \textbf{Severity} & \textbf{Bug Description}                            \\ \midrule
JavaScriptCore & 1  & 301466 & x86-64       & Fixed  & P2       & Major    & OOB Memory Read at function JSC::Wasm::arrayNewElem. \\ \midrule
SpiderMonkey   & 2\textsuperscript{*}  & 1996840 & riscv64      & Fixed & P1       & S2       & Assertion failure: is\_intn(imm, kJumpOffsetBits), at js/src/jit/riscv64/extension/base-riscv-i.h. \\ \bottomrule
\end{tabular}%
}
\end{table*}
\section{Related Works}\label{sec:related-works}

\textbf{Fuzzing}. Fuzzing is a technique for detecting bugs and vulnerabilities in software. Its core idea is to utilize mutation and generation algorithms to produce a large volume of test cases and feed them to the target program, with the goal of identifying unexpected behaviors. Since the advent of AFL ~\cite{AFL}, coverage-guided fuzzing has become the mainstream in both academia and industry and a large body of work has aimed to improve its efficiency. CollAFL~\cite{CollAFL} addresses the hash collision problem of coverage bitmap to obtain precise feedback. AFLFast~\cite{AFLFast}, EcoFuzz~\cite{ecofuzz}, MobFuzz~\cite{MobFuzz}, and other works~\cite{AFL-HIER,K-Scheduler} optimize which seed is selected first for mutation and how many test cases should be generated for each seed according to predefined metrics. MOpt~\cite{MOpt} and DARWIN~\cite{darwin} leverage novel optimization algorithms for scheduling the mutation process, which is similar to the approach adopted in \coolname. GreyOne~\cite{greyone}, VUzzer~\cite{Vuzzer}, and other works~\cite{redqueen, pata} aim to collect more information from the target program by tracking the data flows between input bytes and the operands in comparison instructions and utilize the inferred knowledge to guide mutations and help fuzzers conquer path constraints. In recent years, the large language models (LLMs) and their derivatives have shown remarkable potential in the field of security. Consequently, researchers have also been actively exploring the integration of LLMs with traditional fuzzing practices~\cite{titanfuzz,ChatAFL,MetaMut}. For instance, ChatAFL~\cite{ChatAFL} makes use of LLMs to construct grammars for message types to detect bugs in the protocol implementations and MetaMut~\cite{MetaMut} uses LLMs to generate semantic-aware mutators for testing compilers.

\textbf{JS Fuzzing.} JS is a complex and highly structured kind of input where dedicated fuzzers have been designed. Earlier works~\cite{LangFuzz,Nautilus} generally utilize the context-free grammar to generate new inputs. For example, LangFuzz~\cite{LangFuzz} parses test cases into abstract syntax trees and replaces tree nodes with other nodes of the same type to produce new samples. Subsequent works further focus on passing semantic checks while maintaining grammatical validity. CodeAlchemist~\cite{CodeAlchemist} fragments seed inputs into code bricks and annotates each brick with a set of constraints such as the type of a variable, and these bricks are later combined with all the constraints satisfied. Similarly, Montage~\cite{Montage} operates at code fragment level and trains a neural network model to predict the next fragment given some sequence of fragments, which is later utilized to guide mutations. DIE~\cite{Die} proposes a mutation strategy that preserves interesting structure and type aspects of original regression tests across mutations. Recently, the focus of JS fuzzing has shifted to the JIT compiler. Fuzzilli ~\cite{Fuzzilli} develops a specific intermediate representation FuzzIL and devises a number of generation templates and mutation strategies to trigger JIT. OptFuzz~\cite{OptFuzz} uses an optimization trunk path metric to measure JIT optimization path coverage and guide seed preservation and scheduling. JIT-Picker~\cite{JITPicking} aims to find logic bugs in JIT compilers by comparing the execution hash of interpretation and JIT compilation in the process of fuzzing. FuzzJIT~\cite{FuzzJIT} is also a work that detects JIT compiler logic bugs, which designs a fixed template for triggering JIT compilation by embedding JS code into a for loop.

\textbf{Wasm Fuzzing.} Similar to JS fuzzing, several dedicated fuzzers have been developed for Wasm runtimes. WADIFF \cite{WADIFF} is a differential fuzzer that applies symbolic execution to generate inputs for each instruction. RGFuzz~\cite{RGFuzz} extracts compiler rules from existing Wasm runtimes, which define how to optimize Wasm code, and utilizes them to guide test case generation and find miscompilation bugs. Wasm-smith ~\cite{wasm-smith} is a Wasm module generator which guarantees the static validity of generated modules and can be further combined with other fuzzing frameworks to discover bugs. Waltzz~\cite{Waltzz} proposes the core concept of stack invariant transformation and, based on it, designs several mutators and generator to help fuzzer generate diversified yet valid inputs. WASMaker \cite{WASMaker} is a black-box fuzzer which generates Wasm modules by disassembling and assembling existing Wasm samples, a process similar to that adopted by CodeAlchemist~\cite{CodeAlchemist}.

Although, as demonstrated above, a collection of fuzzing research has been taken separately for JS and Wasm, little attention has been paid to the interaction between these two languages. The \coolname presented in this paper precisely fills this research gap by providing a unified fuzzing framework that systematically explores cross-language behaviors and vulnerabilities emerging from such interactions.
\section{Conclusion}\label{sec:conclusion}

In this paper, we present \coolname, an effective greybox fuzzing framework specifically tailored to discover bugs at the JS-Wasm boundary. To ensure the semantic correctness of the cross-language interactions, \coolname devises a dual-type analysis which effectively infers the JS and Wasm types of each generated variable, which allows our fuzzer to craft valid JS-Wasm interactions in a unified manner. Moreover, \coolname adopts a UCB-1 algorithm to schedule generators and mutators judiciously in the hope of maximizing the code coverage. We have implemented \coolname and evaluated it on three popular JS engines. The evaluation indicates that \coolname has achieved superior code coverage compared to state-of-the-art fuzzers. Currently, \coolname has found two new bugs, one of which is considered security-sensitive and assigned top priority, proving the usefulness of \coolname.
\section*{LLM Usage Considerations}

\noindent LLMs are used for editorial purposes in this paper, and all outputs are inspected by the authors to ensure accuracy and originality.
\section*{Ethics Considerations}

\noindent We have responsibly disclosed all discovered bugs to the relevant vendors, who have confirmed them and undertaken appropriate mitigations.

\bibliographystyle{IEEEtran}
\bibliography{references/ref}
%

\end{document}